\documentclass[a4paper]{article}
\usepackage[normalem]{ulem}
\usepackage{latexsym,amssymb,amsmath,amssymb,xspace,theorem}
\usepackage{epic,eepic}

\theoremstyle{plain} \theoremheaderfont{\scshape}

\newtheorem{Thm}{\bf Theorem}
\newtheorem{Lem}[Thm]{\bf Lemma}

\newtheorem{Cor}[Thm]{ \bf Corollary}
{\theorembodyfont{\rmfamily}

}

\newenvironment{Prf}{{\bf \noindent Proof } }{\hfill$\square$\\}

\newcommand{\ignore}[1]{}

\newcommand{\cqfd}{\unskip\kern 6pt\penalty 500
\raise -2pt\hbox{\vrule\vbox to 10pt{\hrule width 4pt
\vfill\hrule}\vrule}\par}

\newcommand{\dm}{$\delta$-minimum edge-colouring \xspace}
\newcommand{\dms}{$\delta$-minimum edge-colourings \xspace}

\newcommand{\sct}{$\{\alpha,\beta,\gamma\}$}

\input{epsf.sty}
\begin{document}

\title{Tools for parsimonious edge-colouring of graphs with maximum degree three}

\author{J.-L. Fouquet and J.-M. Vanherpe\\
L.I.F.O., Facult\'e des Sciences, B.P. 6759 \\ Universit\'e d'Orl\'eans, 45067 Orl\'eans Cedex 2, FR
}
\date{}
\maketitle

\begin{abstract}
The notion of a $\delta$-minimum edge-colouring was introduced  by J-L. Fouquet (in his french PhD Thesis \cite{FouPhD}). 
Here we present some structural properties of $\delta$-minimum edge-colourings, partially taken from the above thesis. The paper serves as an
auxiliary tool for another paper submitted by the authors to Graphs and Combinatorics.
\end{abstract}

\section{Introduction}
Throughout this note, we shall be concerned with connected graphs with maximum degree $3$. We know by Vizing's  theorem \cite{Viz} that these graphs can be edge-coloured with $4$ colours. 
Let $\phi :\ E(G) \rightarrow \{\alpha,\beta,\gamma,\delta\}$ be a proper edge-colouring of $G$. It is often of interest to try to use one colour (say $\delta$) as few as possible.  In \cite{Fou} we gave without proof
(in French) results on $\delta$-minimum edge-colourings of cubic graphs.  Some of them have been obtained later and independently by Steffen \cite{Ste} and \cite{Ste04}. 
The purpose of Section \ref{section:tools} is to give with their proofs those results as structural properties of $\delta$-minimum edge-colourings.

An edge colouring of $G$ using colours $\alpha,\beta,\gamma,\delta$ is
said to be {\em $\delta$-improper} provided  that adjacent edges having the same colours (if any) are coloured with $\delta$. It is clear that a
proper edge colouring of $G$ is a particular
$\delta$-improper edge colouring.
For a proper or $\delta$-improper edge colouring $\phi$ of $G$ , it will be convenient to denote $E_{\phi}(x)$  ($x \in \{\alpha,\beta,\gamma,\delta\})$ the set of edges coloured with $x$ by $\phi$. 
For $x,y \in\{\alpha,\beta,\gamma,\delta\}$, $x\neq y$, $\phi(x,y)$ is the partial subgraph
of $G$ spanned by these two colours, that is $E_{\phi}(x) \cup E_{\phi}(y)$ (this subgraph being a union of paths and even cycles where the colours $x$ and $y$ alternate). 
Since any two \dms of $G$ have the same number of edges coloured $\delta$ we shall denote by $s(G)$ this number (the {\em colour number} as defined in\cite{Ste}).

 As usual, for any
undirected graph  $G$, we denote by $V(G)$ the set of its vertices
and by $E(G)$ the set of its edges. A {\em strong matching} $C$ in a graph $G$ is a
matching $C$ such that there is no edge of $E(G)$ connecting any two
edges of $C$, or, equivalently, such that $C$ is the edge-set of the
subgraph of $G$ induced on the vertex-set $V(C)$.

\section{Structural properties of $\delta$-minimum edge-colou-rings }\label{section:tools}
The graph $G$ considered in the following series of Lemmas
will have maximum degree $3$.

\begin{Lem} \cite{FouPhD} \label{Lemma:DisjointCyclesIn2Factor} Any
$2$-factor of $G$ contains at least $s(G)$ disjoint odd cycles.
\end{Lem}
\begin{Prf} Assume that we can find a $2$-factor of $G$ with $k<s(G)$ odd cycles. Then let us colour the edges of this $2$-factor with $\alpha$ and $\beta$, except one edge (coloured $\delta$) on each odd cycle of our $2$-factor  and let us colour the remaining edges by $\gamma$. We get hence a new edge colouring $\phi$ with $E_{\phi}(\delta) < s(G)$, impossible.
\end{Prf}

\begin{Lem}  \cite{FouPhD} \label{Lemma:OneVertex2}
Let $\phi$ be a \dm of $G$. Any edge in $E_{\phi}(\delta)$ is
incident to $\alpha$, $\beta$ and $\gamma$. Moreover each such edge
has one end of degree $2$ and the other of degree $3$ or the two
ends of degree $3$.
\end{Lem}
\begin{Prf}
Any edge of $E_{\phi}(\delta)$ is certainly adjacent to $\alpha,\beta$ and $\gamma$. Otherwise this edge could be coloured with the missing colour and we should obtain an edge colouring $\phi'$ with $|E_{\phi'}(\delta)| < |E_{\phi}(\delta)|$.
\end{Prf}

Lemma \ref{Lemma:ImproperDelta} below was proven in \cite{FouVan06Hal}, we give its proof for the sake of completeness.

\begin{Lem} \cite{FouVan06Hal}\label{Lemma:ImproperDelta} Let $\phi$ be a
$\delta$-improper colouring of $G$ then there exists a proper
colouring of $G$ $\phi'$ such that $E_{\phi'}(\delta) \subseteq
E_{\phi}(\delta)$. Moreover, if $\phi$ is improper, then $\phi'$ can be chosen so that $E_{\phi'}(\delta) \subsetneq E_{\phi}(\delta)$.
\end{Lem}
\begin{Prf}
Let $\phi$ be a $\delta$-improper edge colouring of $G$. If $\phi$
is a proper colouring, we are done. Hence, assume that $uv$ and $uw$
are coloured $\delta$. If $d(u)=2$ we can change the colour of $uv$
to $\alpha, \beta$ or $\gamma$ since $v$ is incident to at most two
colours in this set.

If $d(u)=3$ assume that the third edge $uz$ incident to $u$ is also
coloured $\delta$, then we can change the colour of $uv$ for the
same reason as above.

If $uz$ is coloured with $\alpha, \beta$ or $\gamma$, then $v$ and
$w$ are incident to the two remaining colours of the set $\{\alpha,\beta,\gamma\}$ otherwise one of the edges $uv$, $uw$ can
be recoloured with the missing colour. W.l.o.g., consider that $uz$
is coloured $\alpha$ then $v$ and $w$ are incident to $\beta$ and
$\gamma$. Since $u$ has degree $1$ in $\phi(\alpha,\beta)$ let $P$ be the path of $\phi(\alpha,\beta)$ which ends on $u$.   We can assume that $v$ or $w$ (say $v$)
is not the other end vertex of $P$. Exchanging $\alpha$ and $\beta$
along $P$ does not change the colours incident to $v$. But now $uz$
is coloured $\beta$  and we can change the colour of $uv$ to $\alpha$.

In each case, we get hence a new $\delta$-improper edge colouring
$\phi_1$ with $E_{\phi_1}(\delta) \subsetneq E_{\phi}(\delta)$.
Repeating this process leads us to construct a proper edge colouring
of $G$ with $E_{\phi'}(\delta) \subseteq E_{\phi}(\delta)$ as
claimed.
\end{Prf}

\begin{Cor}\label{Cor:DeltaMinimumColoringIsProper}
Any $\delta$-minimum edge-colouring is proper.
\end{Cor}

\begin{Lem} \cite{FouPhD}\label{Lemma:OddCycleAssociated} 
Let $\phi$ be a $\delta$-minimum edge-colouring of $G$. For any edge
$e = uv \in E_{\phi}(\delta)$ with $d(v) \leq d(u)$ there is a colour $x  \in \{\alpha, \beta, \gamma\}$ present at
$v$ and a colour $y \in\{\alpha,\beta,\gamma\}-\{x\}$ present at $u$ such that one of connected
components of $\phi(x, y)$ is a path of even length joining the two ends of $e$.
Moreover, if $d(v)$ = 2, then both colours of $\{\alpha,\beta,\gamma\}-\{x\}$ satisfy the above
assertion.
\end{Lem}
\begin{Prf} Without loss of generality suppose that $x = \gamma$ is present at $v$ and
$\alpha,\beta$ are present at $u$ (see Lemma $2$). Then $u$ is an endvertex of paths in both
$\phi(\alpha,\gamma)$ and $\phi(\beta,\gamma)$, while there is $y \in\{\alpha,\beta\}$ such that $v$ is an endvertex of
a path in $\phi(x,y)$. Without loss of generality assume that both $u$ and $v$ are
endvertices of paths in $\phi(\alpha,\gamma)$. If these paths are disjoint, we exchange the
colours $\alpha$ and $\gamma$ on the path with endvertex $u$ and then recolour $e$ with $\alpha$;
this yields a contradiction to the $\delta$-minimality of $\phi$. To conclude the proof
note that if $d(v) = 2$, then $v$ is an endvertex of paths in both $\phi(\alpha,\gamma)$ and
$\phi(\beta,\gamma)$.
\end{Prf}

 An edge of $E_{\phi}(\delta)$ is in $A_{\phi}$ when its ends
can be connected by a path of $\phi(\alpha,\beta)$, $B_{\phi}$ by a
path of $\phi(\beta,\gamma)$ and $C_{\phi}$   by a path of
$\phi(\alpha,\gamma)$.
From Lemma  \ref{Lemma:OddCycleAssociated} it is clear that if $d(u) = 3$ and $d(v) = 2$ for an edge
$e = uv \in E_{\phi}(\delta)$, the $A_{\phi}$ , $B_{\phi}$ and $C_{\phi}$ are not pairwise disjoint; indeed, if the
colour $\gamma$ is present at the vertex $v$, then $e \in  A_{\phi}\cap B_{\phi}$.

When $e \in A_{\phi}$  we can associate to $e$ the odd cycle
$C_{A_{\phi}}(e)$  obtained by considering the
path of $\phi(\alpha,\beta)$  together with $e$. We define in the
same way $C_{B_{\phi}}(e)$ and $C_{C_{\phi}}(e)$ when $e$ is in
$B_{\phi}$ or $C_{\phi}$.

\begin{Lem}\cite{FouPhD} \label{Lemma:APhiBPhiCPhi} 
If $G$ is a cubic graph then $|A_{\phi}| \equiv |B_{\phi}| \equiv |C_{\phi}| \equiv s(G) \;(mod\; 2)$.
\end{Lem}

\begin{Prf}
$\phi(\alpha,\beta)$ contains $2|A_{\phi}|+|B_{\phi}|+|C_{\phi}|$ vertices of degree $1$ and must be even. Hence we get $|B_{\phi}|\equiv |C_{\phi}| \;(mod\; 2)$. 
In the same way we get $|A_{\phi}|\equiv |B_{\phi}| \;(mod\; 2)$ leading to $|A_{\phi}|\equiv |B_{\phi}| \equiv |C_{\phi}| \equiv s(G) \; (mod\; 2)$.
\end{Prf}
In the following lemma we consider an edge
in $A_{\phi}$, an analogous result  holds true whenever we consider
edges in $B_{\phi}$ or $C_{\phi}$ as well.

\begin{Lem}\cite{FouPhD}\label{Lemma:FondamentalOddCycle}
Let $\phi$ be a \dm of $G$ and let $e$ be an edge in $A_{\phi}$
 then for any edge $e' \in C_{A_{\phi}}(e)$
there is a \dm $\phi'$ such that
$E_{\phi'}(\delta)=E_{\phi}(\delta)-\{e\}\cup \{e'\}$, $e' \in A_{\phi'}$ and
$C_{A_{\phi}}(e)=C_{A_{\phi'}}(e')$. Moreover, each edge outside
$C_{A_{\phi}}(e)$ but incident with this cycle is coloured $\gamma$, $\phi$ and $\phi'$ only differ on the edges of $C_{A_{\phi}}(e)$.
\end{Lem}

\begin{Prf} By exchanging colours $\delta$ and $\alpha$ and
$\delta$ and $\beta$ successively along the cycle $C_{A_{\phi}}(e)$,
we are sure to obtain an edge colouring preserving the number of
edges coloured $\delta$. 
Since we have supposed that $\phi$ is $\delta$-minimum, $\phi$ is proper by Corollary \ref{Cor:DeltaMinimumColoringIsProper}. At each step, the resulting edge colouring remains to be $\delta$-minimum and hence proper.
Hence, there is no edge coloured $\delta$ incident with
$C_{A_{\phi}}(e)$, which means that every such edge is coloured with
$\gamma$.

We can perform these exchanges until $e'$ is coloured $\delta$. In
the \dm $\phi'$ hence obtained, the two ends of $e'$ are joined by a
path of $\phi(\alpha,\beta)$. Which means that $e'$ is in $A_{\phi}$
and $C_{A_{\phi}}(e)=C_{A_{\phi'}}(e')$.
\end{Prf}

For each edge $e \in E_{\phi}(\delta)$ (where $\phi$ is a \dm of
$G$) we can associate one or two odd cycles following the fact
that $e$ is in one or two sets from among $A_{\phi}$, $B_{\phi}$,
$C_{\phi}$. Let ${\mathcal C}$ be the set of odd cycles associated
to edges in $E_{\phi}(\delta)$.

\begin{Lem} \cite{FouPhD}\label{Lemma:IsolatedVertex2}
 For each cycle  $C \in {\mathcal C}$,
there are no two consecutive vertices with degree two.
\end{Lem}
\begin{Prf}
Otherwise, we exchange colours along $C$ in order to put the colour $\delta$  on the corresponding edge and, by Lemma \ref{Lemma:OneVertex2}, this is impossible in a \dm.
\end{Prf}

\begin{Lem}\cite{FouPhD} \label{Lemma:DisjointOddCycles}
Let $e_1,e_2 \in E_{\phi}(\delta)$, such that $e_1\neq e_2$  and let $C_1,C_2 \in \mathcal C$
be such that $C_1\neq C_2$, $e_1 \in E(C_1)$ and $e_2 \in E(C_2)$
then $C_1$ and $C_2$ are (vertex) disjoint.
\end{Lem}
\begin{Prf}  If $e_1$ and $e_2$ are contained in the same set $A_{\phi}$, $B_{\phi}$, or $C_{\phi}$, we are done since their respective ends are joined by an alternating path of $\phi(x,y)$ 
for some two colours $x$ and $y$ in \sct.

Without loss of generality assume that $e_1 \in A_{\phi}$ and $e_2 \in B_{\phi}$. Assume moreover that there exists an edge $e$ such that $e \in C_1 \cap C_2$. 
We have hence an edge $f \in C_1$ with exactly one end on $C_2$. We can exchange colours on $C_1$ in order to put the colour $\delta$ on $f$, which is impossible by Lemma \ref{Lemma:FondamentalOddCycle}.
\end{Prf}

\begin{Lem}\cite{FouPhD}\label{Lemma:OneVertexInNeighborhood}
Let $e_1=uv_1$ be an edge of $E_{\phi}(\delta)$ such that $v_1$ has
degree $2$ in $G$. Then $v_1$ is the only vertex in $N(u)$ of degree
 $2$ and for any other edge $e_2 \in E_{\phi}(\delta)$,  $\{e_1,e_2\}$ induces a $2K_2$.
\end{Lem}
\begin{Prf}
We have seen in Lemma \ref{Lemma:OneVertex2} that $uv_1$ has one end of degree $3$ while the other has degree $2$ or $3$. Hence, we have $d(u)=3$ and $d(v_1)=2$. 
Let $v_2$ and $v_3$ the other neighbours of $u$.
We can suppose without loss of generality that $uv_2$ is coloured $\alpha$, $uv_3$ is coloured $\beta$ and, finally $v_1$ is incident to an edge coloured $\gamma$, say $v_1v_4$.

Assume first that $d(v_2)=2$. An alternating path of $\phi(\beta,\gamma)$ using the edge $uv_3$ ends with the vertex $v_1$, moreover $v_2$ is incident to an edge coloured $\gamma$ since an 
alternating path of $\phi(\alpha,\gamma)$ using the edge 
$uv_2$ ends with $v_1$ (see Lemma \ref{Lemma:OddCycleAssociated}), then, 
exchanging the colours along the component of $\phi(\beta,\gamma)$ containing $v_2$ allows us to colour $uv_2$ with $\gamma$ and $uv_1$ with $\alpha$.
 A new edge colouring $\phi'$ so obtained is such that $|E_{\phi'}(\delta)| \leq |E_{\phi}(\delta)|-1$, impossible.

Thus, $d(v_2 ) = 3$, and, by symmetry, $d(v_3 ) = 3$. We know that $e_1\in B_{\phi}\cap C_{\phi}$  (see Lemma \ref{Lemma:OddCycleAssociated}). 
By Lemma \ref{Lemma:FondamentalOddCycle}, since $e_1\in C_{B_{\phi}}(e_1)$ the edges incident to $v_3$ in $C_{B_{\phi}}(e_1)$ are coloured with $\beta$ and $\gamma$ and 
the third edge incident to $v_3$ is coloured $\alpha$. Similarly the vertex $v_2$ being on $C_{B_{\phi}}(e_1)$ is incident to colours $\alpha$ and $\gamma$ 
and the third edge incident to $v_2$ is  coloured $\beta$. Moreover the vertex $v_4$ being on both cycles $C_{B_{\phi}}(e_1)$ and $C_{C_{\phi}}(e_1)$ is incident to colours $\alpha$, $\beta$,
 $\gamma$. Hence no edge coloured $\delta$ can be incident to $v_2$ nor $v_3$ nor $v_4$. 
It follows that for any edge $e_2$ in $E_{\phi}(\delta)-\{e_1\}$, the set $\{e_1,e_2\}$ induces a $2K_2$.
\end{Prf}
\begin{Lem}\cite{FouPhD}\label{Lemma:2K2+SinonAuPlusUneArete}
Let $e_1$ and $e_2$ be two edges of $E_{\phi}(\delta)$  $e_1\neq e_2$.
If $e_1$ and $e_2$ are contained in two distinct sets of $A_{\phi},B_{\phi}$, $C_{\phi}$ then $\{e_1,e_2\}$ induces a $2K_2$, otherwise $e_1,e_2$ are joined by at most one edge.
\end{Lem}

\begin{Prf} By Lemma \ref{Lemma:OneVertexInNeighborhood} one can suppose that all vertices incident with $e_1$, $e_2$ are of degree $3$ so that there is exactly one of the sets 
$A_{\phi}$, $B_{\phi}$ or $C_{\phi}$ containing $e_1$ (respectively $e_2$).

Assume in the first stage that $e_1 \in A_{\phi}$ and $e_2 \in B_{\phi}$. Since $C_{A_{\phi}}(e_1)$ and $C_{B_{\phi}}(e_2)$ are disjoint by Lemma \ref{Lemma:DisjointOddCycles}, 
we know that $e_1$ and $e_2$ have no common vertex. The edges having exactly one end in $C_{A_{\phi}}(e_1)$ are coloured $\gamma$ while those having exactly one end 
in $C_{B_{\phi}}(e)$ are coloured  $\alpha$. Hence there is no edge between $e_1$ and $e_2$ as claimed.

Assume in the second stage that $e_1=u_1v_1, e_2=u_2v_2 \in A_{\phi}$. Since $C_{A_{\phi}}(e_1)$ and $C_{B_{\phi}}(e_2)$ are disjoint by Lemma
\ref{Lemma:DisjointOddCycles}, we can consider that $e_1$ and $e_2$ have no common vertex. 
The edges having exactly one end in $C_{A_{\phi}}(e_1)$ or $C_{B_{\phi}}(e_2)$ are coloured $\gamma$. Assume that $u_1u_2$ and $v_1v_2$ are edges of $G$. 
We may suppose  without loss of generality that $u_1$ and $u_2$ are incident to $\alpha$ while  $v_1$ and $v_2$ are incident to $\beta$ (if necessary, colours $\alpha$ and $\beta$ can be exchanded 
 on $C_{A_{\phi}}(e_1)$ and $C_{B_{\phi}}(e_2)$).  We know that $u_1u_2$ and $v_1v_2$ are coloured $\gamma$. 
Let us colour $e_1$ and $e_2$ with $\gamma$ and $u_1u_2$ with $\beta$ and $v_1v_2$ with $\alpha$. 
We get a new edge colouring $\phi'$ where $|E_{\phi'}(\delta)| \leq |E_{\phi}(\delta)|-2$, contradiction since $\phi$ is a \dm.
\end{Prf}

\begin{Lem} \cite{FouPhD}\label{Lemma:Oneedgewiththree}
Let $e_1, e_2$ and $e_3$ be three distinct edges of
$E_{\phi}(\delta)$ contained in the same  set $A_{\phi},B_{\phi}$ or
$C_{\phi}$. Then $\{e_1,e_2,e_3\}$  induces a subgraph with at most
four edges.
\end{Lem}
\begin{Prf} Without loss of generality assume that $e_1=u_1v_1, e_2=u_2v_2$ and $ e_3=u_3v_3\in A_{\phi}$. From Lemma \ref{Lemma:2K2+SinonAuPlusUneArete} we have just to suppose that (up to the names of vertices) $u_1u_3 \in E(G)$ and $v_1v_2 \in E(G)$. Possibly, by exchanging  the colours $\alpha$ and $\beta$ along the $3$ disjoint paths of $\phi(\alpha,\beta)$ joining the ends of each edge $e_1,e_2$ and $e_3$, we can suppose that $u_1$ and $u_3$ are incident to $\beta$ while $v_1$ and $v_2$ are incident to $\alpha$. Let $\phi'$ be obtained from $\phi$ when $u_1u_3$ is coloured with $\alpha$, $v_1v_2$ with $\beta$ and $u_1v_1$ with $\gamma$. It is easy to check that $\phi'$ is a proper edge-colouring with $|E_{\phi'}(\delta)| \leq |E_{\phi}(\delta)|-1$, contradiction since $\phi$ is a \dm.
\end{Prf}
Let us summarize most of the above results in a single Theorem.
\begin{Thm}\label{Thm:Summarize}
Let $G$ be a graph of maximum degree $3$ and $\phi$ be a $\delta$-minimum colouring of $G$. Then the following hold.\\
\begin{enumerate}
\item $E_{\phi}(\delta)=A_{\phi}\cup B_{\phi}\cup C_{\phi}$ where an edge $e$ in $A_{\phi}$ ($B_{\phi}$, $C_{\phi}$ respectively) belongs to a uniquely determined cycle
$C_{A_{\phi}}(e)$ ($C_{B_{\phi}}(e)$, $C_{C_{\phi}}(e)$ respectively) with precisely one edge coloured $\delta$ and the other edges being alternately coloured $\alpha$ and 
$\beta$ ($\beta$ and $\gamma$, $\alpha$ and $\gamma$ respectively). 
\item Each edge having exactly one vertex in common with some edge in $A_{\phi}$ ($B_{\phi}$, $C_{\phi}$ respectively) is coloured $\gamma$ ($\alpha$, $\beta$, respectively).
\item The multiset of colours of edges of $C_{A_{\phi}}(e)$ ($C_{B_{\phi}}(e)$, $C_{C_{\phi}}(e)$ respectively) can be permuted to obtain a (proper) $\delta$-minimum edge-colouring of G in which
the colour $\delta$ is moved from $e$ to an arbitrarily prescribed edge.
\item No two consecutive vertices of $C_{A_{\phi}}(e)$ ($C_{B_{\phi}}(e)$, $C_{C_{\phi}}(e)$ respectively) have degree 2.
\item The cycles from $1$ that correspond to distinct edges of $E_{\phi}(\delta)$ are vertex-disjoint.
\item If the edges $e_1, e_2, e_3 \in E_{\phi}(\delta)$ all belong to $A_{\phi}$ ($B_{\phi}$, $C_{\phi}$ respectively), then the set $\{e_1 , e_2 , e_3 \}$ 
induces in $G$ a subgraph with at most $4$ edges.
\end{enumerate}
\end{Thm}

\bibliographystyle{plain}

\bibliography{ToolsForParcimonious}

\end{document}